\documentclass[aps,twocolumn,showpacs,prl]{revtex4}
\usepackage[dvips]{graphicx}
\usepackage[english]{babel}
\usepackage[T1]{fontenc}
\usepackage{mathrsfs}
\usepackage[tbtags]{amsmath}
\usepackage{amssymb}
\usepackage{amsxtra}
\usepackage{amsopn}
\usepackage{latexsym}
\usepackage[mathcal]{eucal}

\newcommand{\BE}{\begin{equation}}
\newcommand{\EE}{\end{equation}}
\newcommand{\BA}{\begin{align}}
\newcommand{\EA}{\end{align}}

\newcommand{\nn}{\nonumber}

\begin{document}

\title{Perturbation theory of non-perturbative QCD}

\author{Fabio Siringo}

\affiliation{Dipartimento di Fisica e Astronomia 
dell'Universit\`a di Catania,\\ 
INFN Sezione di Catania,
Via S.Sofia 64, I-95123 Catania, Italy}

\date{\today}
\begin{abstract}
Perturbation theory is shown to be working in the IR limit of
pure SU(3) Yang-Mills theory in Landau gauge by an unconventional setting of the perturbative expansion.
A dynamical mass is predicted for the gluon and the lattice data are reproduced fairly well by a second-order expansion, 
without any free parameter. The effective running coupling is small in the IR and the approximation
can be improved by inclusion of higher order terms.
\end{abstract}



\maketitle

It is well known that perturbation theory (PT) breaks down in the infrared (IR) limit
of QCD. The coupling grows up and eventually diverges at the low-energy scale $\Lambda_{QCD}\approx$ 200 MeV,
so that our knowledge of the very important low-energy phenomenology, including hadron masses and quark confinement,
must rely on non-perturbative methods and mainly on numerical lattice simulations. 
In the IR, we still miss a full analytical description that could be regarded
as sound and reliable as PT is in the ultraviolet (UV).
On the lattice, by non-perturbative renormalization schemes like moment subtraction (MOM), the renormalized coupling
turns out to be finite and not too large in the IR, so that its divergence must be regarded as an artifact of
standard PT. A finite coupling would suggest that a perturbative approximation could be set up and that the failure
of standard PT is a consequence of the usual perturbative approach more than a problem with the general perturbative
method by itself.

In this paper, by an unconventional setting of the perturbative method, PT is shown to be working in the IR limit of
pure SU(3) Yang-Mills (YM)  in Landau gauge, 
yielding a first-principle analytical description that reproduces the lattice
data fairly well, without any free parameter that was not in the original Lagrangian. Moreover, the approximation
can be improved, order by order, by inclusion of higher order terms, as for any other perturbative expansion.

A key role is played by the dynamical mass that the gluon acquires in non-perturbative approximations, as its
appearence is prohibited by gauge invariance at any order of standard PT. That is a signature of an intrinsic
failure of the usual perturbative expansion in the IR 
and the gluon mass can be regarded as a pure non-perturbative effect.
The success of any perturbative expansion relies on the choice of the zeroth order approximation that should
capture most of the physics, leaving minor details to be defined by inclusion of the interaction. 
While a free massless gluon is a good starting point in the UV, no finite mass can arise in PT if it is not 
already present in the zeroth order of the expansion. On the other hand, it has been recently shown that, by inclusion
of a mass term in the Lagrangian, PT turns out to be perfectly viable in the IR, yielding one-loop results that
fit the lattice data very well\cite{tissier10,tissier11}. Thus the gluon mass seems to absorb most of the
non-perturbative effects and having included the mass in the zeroth order propagator, the effects of 
the interaction can be treated by standard PT. Moreover, in the UV limit the mass becomes irrelevant, and the
modified Lagrangian maintains the correct high energy limit.

However, inclusion of a mass by hand would not be a satisfactory solution of the problem for many reasons.
First of all, the mass term is forbidden by gauge invariance, together with any counterterm that should be
required for its renormalization. In addition, while a free mass-parameter would be useful for interpolating
experimental and lattice data, it would be an extra parameter that was not in the original Lagrangian, yielding a
phenomenological model at most.

The choice of a more clever zeroth order Lagrangian does not require the inclusion of extra terms because we have the
freedom of splitting the original action in any two parts, yielding different perturbative expansions.
In fact, that freedom suggests a way to optimize the expansion by a variational evaluation of the best zeroth
order action\cite{gep2,varqed,varqcd,genself}. In other words, we may add a mass term to the zeroth order Lagrangian
and subtract the same term in the interaction, without affecting the total Lagrangian that maintains its gauge
invariance. While the total action does not depend on that mass, any finite-order expansion would depend on
the mass parameter that can be optimized by some variational ansatz in order to get the best perturbative expansion.
The optimized PT is not gauge invariant at any finite order, but gauge invariance must be recovered approximately in
the physical observables if they converge, order by order, towards the exact ones.

For a pure $SU(N)$ YM theory in Landau gauge, dropping color indices, the zeroth-order gluon propagator can be taken
as
\BE
\Delta^{\mu\nu} (p)=\left[g^{\mu\nu}  - \frac{p^\mu p^\nu}{p^2}\right] \Delta_0(p)
\label{D}
\EE
where $\Delta_0$ is the propagator of a massive free gluon
\BE
\Delta_0 (p)^{-1}= -p^2+m^2
\EE
while the zeroth-order ghost propagator $G_0$ remains unchanged (massless).
By standard Dyson equations, the dressed gluon and ghost propagators read
\BE
\Delta (p)^{-1}= -p^2+m^2-\Pi^\star(p)
\label{Delta}
\EE
\BE
G(p)^{-1}=p^2-\Sigma^\star (p)
\EE
where $\Pi^\star$ is the proper polarization and $\Sigma^\star$ is the proper self energy.
These functions are to be evaluated, order by order, by PT as expansions in powers of the
total interaction, including a mass counter-term that cancels the mass in the total Lagrangian.
Namely, the total interaction ${\cal L}_{tot}$ can be written as
\BE
{\cal L}_{tot}=\delta\left[-\frac{1}{2} m^2 A^\mu A_\mu+{\cal L}_{int}\right]
\EE
where $\delta$ is a power-counting parameter to be set to one at the end of the calculation
and ${\cal L}_{int}$ is the original interaction of YM theory, including the ghost vertex and
three- and four-gluon vertices that contain powers of the bare YM coupling $g$.
Thus, the perturbative expansion is not a loop-wise expansion in powers of the coupling $g$, 
but must be written as a $\delta$-expansion where graphs with a different number of loops may
coexist at the same order\cite{stancu2}. In Fig.1 the one-particle irredicible graphs are displayed 
up to second order in $\delta$.

\begin{figure}[t] \label{fig:graph}
\centering
\includegraphics[width=0.45\textwidth,angle=0]{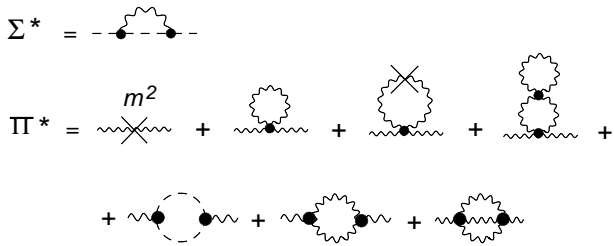}
\caption{First and second order graphs contributing to the ghost self energy and the gluon
proper polarization. The filled circles are the bare vertices of ${\cal L}_{int}$ while the cross is
the mass counter-term $m^2$.}
\end{figure}

Of course the proper functions $\Sigma^\star$, $\Pi^\star$ do contain divergences and a regularization
scheme is in order. While dimensional regularization is suited for standard loop-wise expansions where
gauge invariance is preserved at any finite order, here gauge invariance is already lost and
a more physical non-perturbative scheme would be preferred, like a simple cutoff. 
Restricting all integrals
to $p^2<\Lambda^2$ in the Euclidean space, the graphs in Fig.1 can be evaluated for any choice of the
bare YM coupling $g$ and for any value of the parameter $m^2$. 
Moreover we can make contact with lattice
simulations where a natural cutoff is provided by the lattice spacing and we can use the same non-perturbative
{\it multiplicative} renormalization scheme, thus avoiding any insertion of counter-terms in the
Lagrangian.  As for a lattice, we assume that the bare coupling  $g=g(\Lambda)$ is a function of the cutoff
and the physical content of the theory is invariant for a change of scale $\Lambda\to \Lambda^\prime$ followed by
a change of coupling $g\to g^\prime$. Then,
physical renormalized functions that do not depend on the cutoff nor on the coupling 
can be defined at an arbitrary scale $\mu$ by the usual multiplicative renormalization. 
That is equivalent to say that all bare functions, at different couplings, 
should go one on top of the other by scaling, yielding a test for the accuracy of the perturbative approximation.

A drawback of the simple choice of a cutoff is the occurrence of spurious quadratic divergences besides the usual
logarithmic terms that are observed on the lattice. These quadratic terms must cancel in the exact propagators
and polarization functions so that their appearance can be regarded as an artifact of the finite-order
approximation. Moreover, while the exact functions cannot depend on $m^2$, the finite-order expansion does
depend on the choice of the mass parameter and the spurious divergences are sensitive to the choice of $m^2$.
That would suggest a simple criterion for determining the best mass parameter: we assume that the best
expansion is the one where the quadratic divercences are vanishing (or smaller).
At any finite order, that choice would give an expansion free of spurious divergences and no free parameters
at all, allowing for a challanging comparison of its predictions with lattice data.
Besides its formal elegance, that choice is quite resonable if one wants to recover the correct scaling
properties of the theory.
 
An other similar choice could rely on the use of Stevenson's principle of minimal sensitivity\cite{minimal} that
would require $m^2$ to make the physical observables stationary. In fact, while in principle $m^2$ can be
chosen at will, any sensible physical prediction should not depend on its choice. The mass parameter $m^2$
is not a physical observable by itself, and different strategies for its optimization may lead to very similar
physical results if $m^2$ is close to a stationary point where small changes in the mass of the zeroth order
propagator are compensated by the changes on the interaction.

It is not difficult to show that the second-order proper functions in Fig.1 can be made free of quadratic
divergences by a special choice of the mass parameter. We may neglect the last graph of Fig.1, namely the
two-loop sunrise graph, as it has been shown to be very small and basically constant\cite{genself}.
Its inclusion would just make the derivation awkward without any substantial change in the result.
Moreover, we may neglect subleading terms that vanish in the limit $\Lambda\to\infty$. The quadratic
divergences can be extracted by the polarization function at $p=0$ as the difference $\Pi^\star(p)-\Pi^\star(0)$
contains logarithmic terms at most. By an explicit calculation of the graphs in Fig.1 we find\cite{genself}
\begin{align}
\Pi^\star(0)&= m^2\left(1+\frac{5}{3}a\right)
+m^2 a\left(a-\frac{4}{3}\right)\log\left(1+\frac{\Lambda^2}{m^2}\right)\nn\\
&-m^2 a^2 \log^2 \left(1+\frac{\Lambda^2}{m^2}\right)\nn\\
&+a\Lambda^2\left[a\log\left(1+\frac{\Lambda^2}{m^2}\right)-\frac{4}{9}-a\right]
\label{pol0}
\end{align}
where the effective bare coupling $a=9N\alpha/(16\pi)$ and $\alpha=g^2/(4\pi)$. For $N=3$ the effective
coupling $a\approx 0.04 g^2$ is quite small even when $g$ is rather large.
We recognize a term $(a^2\log^2)$  arising from the two-loop tadpole: when $\Lambda\gg m$ the infinite
series of tadpoles would eventually require a resummation of the large logarithms.

The quadratic divergence cancel exactly if the mass parameter is taken as
\BE
\left(\frac{m^2}{\Lambda^2}\right)=\left[\exp\left(1+\frac{4}{9a}\right)-1\right]^{-1}
\label{mass}
\EE
which has the evident non-perturbative behaviour $m^2\sim \exp(-{\rm const.}/a)$ in the limit $a\to 0$.
Inserting Eq.(\ref{mass}) back in Eq.(\ref{pol0}) the term $(a^2 \log^2)$  cancels exactly, yielding the
simple linear expression
\BE
\Pi^\star(0)= m^2\left[1+\frac{5}{3}a-\frac{16}{9}a\log\left(1+\frac{\Lambda^2}{m^2}\right)\right].
\EE
By insertion of Eq.(\ref{mass}) again, we can make the logarithmic divergence implicit through the coupling and
write 
\BE
\Pi^\star(p)= m^2\left[1-\frac{64}{81}-\frac{1}{9}a\right]+\left[\Pi^\star(p)-\Pi^\star(0)\right]
\label{polM}
\EE
where the last term $\Pi^\star(p)-\Pi^\star(0)$ is just given by the sum of gluon and ghost one-loop sunrise graphs,
the only ones that depend on $p$. By Eq.(\ref{Delta}) the dressed propagator acquires the dynamical mass 
$[m^2-\Pi^\star(p)]$ that in the limit $p\to 0$ defines a {\it physical} mass
\BE
M^2=m^2\left(\frac{64}{81}+\frac{1}{9}a\right)
\label{Mass}
\EE
which survives after the cancellation of the mass terms in the zeroth order propagator and in the interaction.
After renormalization, the mass $M^2$ is expected to be invariant for any change of coupling and cutoff. 

In order to compare with the phenomenological model of Ref.\cite{tissier11} we explore the optimized
second-order expansion for $N=3$ in the large-coupling range $g=4-5$ where that model was found in perfect 
agreement with
the lattice data. We find a good agreement with that model, which is not surpising as Eq.(\ref{polM}) would be
equivalent to the result of a one-loop calculation with a phenomenological mass inserted by hand 
and set equal to $M^2$.  

\begin{figure}[t] \label{fig:Log}
\centering
\includegraphics[width=0.35\textwidth,angle=-90]{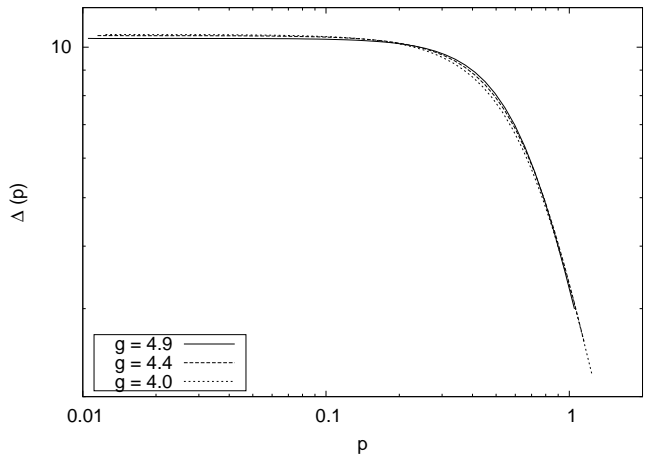}
\caption{Log-Log plot of the renormalized gluon propagator for $N=3$ in arbitrary units for $g=$ 4, 4.4 and 4.9.}
\end{figure}

\begin{figure}[b] \label{fig:Delta}
\centering
\includegraphics[width=0.35\textwidth,angle=-90]{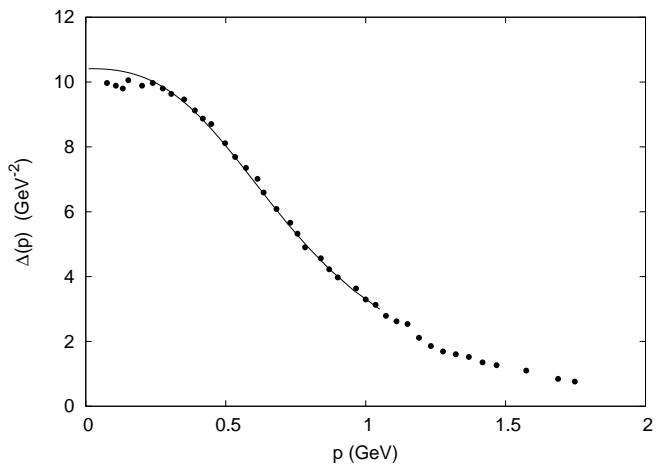}
\caption{The renormalized gluon propagator in physical units ($\Lambda=1.05$ Gev) for $N=3$ and $g=4.9$. 
The points are the lattice data of Ref.\cite{bogolubsky} for $\beta=5.7$ and $L=96$.}
\end{figure}

First of all, in Fig.2 we check that the gluon propagators can be renormalized by a multiplicative scale factor
and  translated one on top of the other by a translation in a log-log plot. As one would expect 
by a finite-order approximation, the scaling properties are not stisfied exactly 
but we can still extract a renormalized propagator which is almost independent of the coupling. The residual
dependence suggests that the agreement with lattice data would be at its best for a special value of the 
coupling that turns out to be $g\approx 4.9$ in the phenomenological model of Ref.\cite{tissier11}.
That small dependence should decrease by inclusion of higher order terms.

\begin{figure}[t] \label{fig:ghostdress}
\centering
\includegraphics[width=0.35\textwidth,angle=-90]{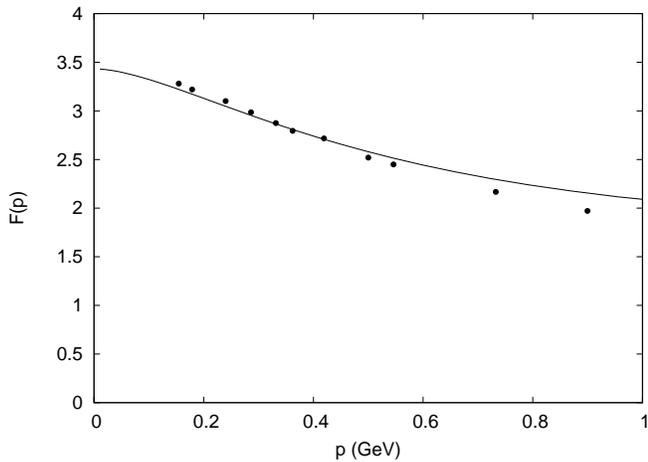}
\caption{The renormalized ghost dressing function $F(p)=p^2G(p)$ for $N=3$ and $g=4.9$
in the same physical units of Fig.3 ($\Lambda=1.05$ Gev). 
The points are the lattice data of Ref.\cite{bogolubsky} for $\beta=5.7$ and $L=96$.}
\end{figure}

\begin{figure}[b] \label{fig:alp}
\centering
\includegraphics[width=0.35\textwidth,angle=-90]{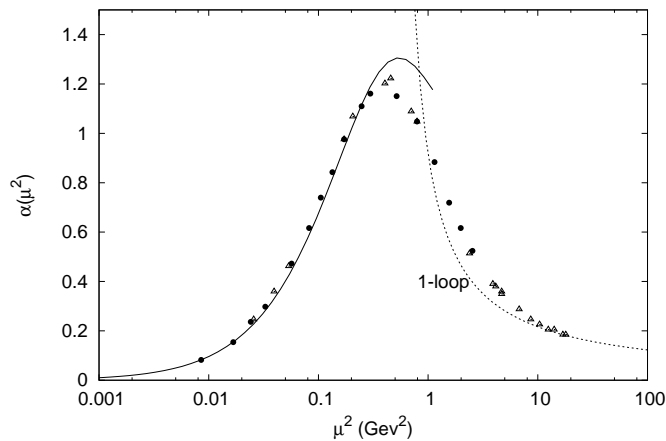}
\caption{The running coupling as defined in Eq.(\ref{alpha}), for $N=3$ and $g=4.9$, with the same renormalization
constants of Fig.3 and Fig.4 (solid line). 
The points are the lattice data of Ref.\cite{bogolubsky} for $\beta=5.7$, $L=80$ (circles) 
and $L=64$ (triangles). The standard PT one-loop behaviour is reported as a dotted line.}
\end{figure}

Having no experimental data on pure SU(3) we fix the physical energy scale by a comparison with the lattice data
of Ref.\cite{bogolubsky} and renormalize the propagator at the same energy of that work. The cutoff turns out to
be $\Lambda=1.05$ Gev at $g=4.9$. That value is small but not too far from the inverse lattice spacing that is
$1/\ell\approx 1.16$ GeV in the simulation\cite{bogolubsky}. The renormalized gluon propagator is shown in Fig.3
together with the lattice data. The agreement is very good for a first-principle second order calculation with
no free parameters.
The physical mass from Eq.(\ref{Mass}) turns out to be $M=0.56$ GeV and is constant in the range $g=4-4.9$ as
shown in table I. That value is not far from the phenomenological parameter $m=0.54$ GeV extracted by a best fit of
data in Ref.\cite{tissier11}. 

\begin{table}[!t]
\centering 
\begin{tabular}{c c c c c} 
\hline\hline 
$g$ &\qquad $m^2/\Lambda^2$ \quad &\qquad $\Lambda$ (GeV)\quad &\qquad $m$ (GeV)\quad & \qquad $M$ (GeV) \quad \\
[0.5ex] 
\hline 
4.0  & 0.238 & 1.24 & 0.60 & 0.56  \\ 
4.4  & 0.274 & 1.15 & 0.60 & 0.56  \\ 
4.9  & 0.313 & 1.05 & 0.59 & 0.56  \\ 
\hline 
\end{tabular}
\label{table} 
\caption{Masses for $N=3$ at the energy scale of Ref.\cite{bogolubsky}.}
\end{table}

In Fig.4 the renormalized ghost dressing function $F(p)=p^2G(p)$ is 
reported with the same energy units ($\Lambda=1.05$ Gev) and compared with the lattice data. The agreement
is fairly good but less remarkable than for the gluon, with a deviation that increases as $p$ approaches the cutoff. 
Since it is well known that a fixed-order calculation needs renormalization group (RG) corrections in the UV limit,
we do not bother about the large-energy predictions of the model. As largely discussed in Ref.\cite{tissier11},
the mass parameter becomes irrelevant in the UV and the optimized perturbative expansion is equivalent to
the standard PT in that limit. On the other hand, in the IR the optimized perturbative expansion 
open the way to a first-principle
perturbative description of the phenomenology that can be improved order by order and reach a degree of accuracy
comparable to the UV limit. In fact, by a MOM-Taylor scheme\cite{taylor} with the vertex renormalization constant 
set to one in Landau gauge, a running coupling can be defined by the RG invariant quantity
\BE
\alpha(\mu^2)=\frac{g^2}{4\pi}\Delta(\mu) F^2(\mu)
\label{alpha}
\EE
that has been measured on the lattice in Ref.\cite{bogolubsky}. The coupling dose not get too large in the IR
and can be evaluated in PT, order by order, from first principles, without any information from the lattice.
In fact, renormalizing at the same scale of Figs. 3 and 4, we can evaluate the
running coupling directly from Eq.(\ref{alpha}) and compare the result with the lattice data. As shown in Fig.5
the agreement is very good in the IR. 

These findings are encouraging and we expect that eventually, 
including quarks in the formalism would lead to a full perturbative description of QCD in the IR.

\end{document}